\documentclass[sigconf, authorversion]{acmart}
\usepackage[utf8]{inputenc}
\usepackage{mathrsfs}
\usepackage{multirow}
\usepackage{graphicx}
\usepackage{textcomp}
\usepackage{amsthm}
\usepackage{subfig}
\usepackage{graphics}
\usepackage{hyperref}
\usepackage[font=small,labelfont=bf]{caption}
\usepackage{algorithm}
\usepackage[noend]{algpseudocode}
\usepackage{balance} 

\newcommand{\model}[0]{\textsc{CS-MLGCN }}
\newcommand{\V}[0]{\mathcal{V}}
\newcommand{\E}[0]{\mathcal{E}}
\newcommand{\GG}[0]{\mathcal{G}}
\newcommand{\X}[0]{\mathcal{X}}
\newcommand{\CC}[0]{\mathscr{C}}
\newcommand{\M}[0]{\mathcal{M}}
\newcommand{\cvec}[0]{\mathbf{c}}

\newcommand{\vencoder}[0]{\textit{View Encoder }}
\newcommand{\qencoder}[0]{\textit{Query Encoder }}
\newcommand{\bb}[0]{\mathbf{b}}
\newcommand{\W}[0]{\mathbf{W}}
\newcommand{\Dr}[0]{\mathrm{Dr}}

\AtBeginDocument{%
  \providecommand\BibTeX{{%
    \normalfont B\kern-0.5em{\scshape i\kern-0.25em b}\kern-0.8em\TeX}}}

\newcommand{\head}[1]{\vspace{1.7mm}\noindent{{\bf #1.}}}

\newtheoremstyle{sig}
  {}
  {}
  {\itshape}
  {}
  {\scshape}
  {.}
  {.5em}
  {#1 #2\thmnote{\quad(#3)}}

\theoremstyle{sig}

\newtheorem{problem}{Problem}

\copyrightyear{2022}
\acmYear{2022}
\setcopyright{acmcopyright}\acmConference[CIKM '22]{Proceedings of the 31st ACM International Conference on Information and Knowledge Management}{October 17--21, 2022}{Atlanta, GA, USA}
\acmBooktitle{Proceedings of the 31st ACM International Conference on Information and Knowledge Management (CIKM '22), October 17--21, 2022, Atlanta, GA, USA}
\acmPrice{15.00}
\acmDOI{10.1145/3511808.3557572}
\acmISBN{978-1-4503-9236-5/22/10}

\begin{document}

\setlength{\textfloatsep}{5pt}
\everypar{\looseness=-1}

\title{\model : Multiplex Graph Convolutional Networks for Community Search in Multiplex Networks}

\author{Ali Behrouz}
% \authornote{\relax}
\authornote{$\:$These authors contributed equally.}
\affiliation{%
  \institution{University of British Columbia}
  \city{Vancouver}
  \state{BC}
  \country{Canada}
}
\email{alibez@cs.ubc.ca}

\author{Farnoosh Hashemi}  
\authornotemark[1]
\affiliation{%
  \institution{University of British Columbia}
  \city{Vancouver}
  \state{BC}
  \country{Canada}
}
\email{farsh@cs.ubc.ca}

\begin{abstract}
Community Search (CS) is one of the fundamental tasks in network science and has attracted much attention due to its ability to discover personalized communities with a wide range of applications. Given any query nodes, CS seeks to find a densely connected subgraph containing query nodes. Most existing approaches usually study networks with a single type of proximity between nodes, which defines a single view of a network. However, in many applications such as biological, social, and transportation networks, interactions between objects span multiple aspects, yielding networks with multiple views, called multiplex networks. Existing CS approaches in multiplex networks adopt pre-defined subgraph patterns to model the communities, which cannot find communities that do not have such pre-defined patterns in real-world networks. In this paper, we propose a query-driven graph convolutional network in multiplex networks, \model\!, that can capture flexible community structures by learning from the ground-truth communities in a data-driven fashion. \model first combines the local query-dependent structure and global graph embedding in each type of proximity and then uses an attention mechanism to incorporate information on different types of relations. Experiments on real-world graphs with ground-truth communities validate the quality of the solutions we obtain and the efficiency of our model.
\end{abstract}

\begin{CCSXML}
<ccs2012>
   <concept>
       <concept_id>10010147.10010257.10010293.10010294</concept_id>
       <concept_desc>Computing methodologies~Neural networks</concept_desc>
       <concept_significance>300</concept_significance>
       </concept>
   <concept>
       <concept_id>10002950.10003624.10003633.10010917</concept_id>
       <concept_desc>Mathematics of computing~Graph algorithms</concept_desc>
       <concept_significance>300</concept_significance>
       </concept>
   <concept>
       <concept_id>10002951.10003260.10003282.10003292</concept_id>
       <concept_desc>Information systems~Social networks</concept_desc>
       <concept_significance>300</concept_significance>
       </concept>
 </ccs2012>
\end{CCSXML}

% \ccsdesc[300]{Information systems~Social networks}


\maketitle
% \vspace{-1ex}
\section{Introduction}\label{sec:introduction}
Identifying communities is a fundamental problem in network science and has been traditionally addressed with the aim of determining an organization of a given network into subgraphs that express densely connected groups of nodes~\cite{Community_Detection_main}. Given any query nodes, community search (CS) aims to find communities covering the query nodes~\cite{k-core-community}. Recently, CS has attracted much attention due to its ability to discover personalized communities with a wide range of applications, e.g., identifying functional systems in brain networks~\cite{FirmTruss, ML-random-walk}, classifying brain networks~\cite{FirmTruss}, protein complex identification~\cite{attribute-community}, and team formation~\cite{TeamFormation}. 

While significant research effort has been devoted to the study of CS, 
the majority of works assume there only exists a single type of proximity between nodes in a network, whereas, in applications such as social, brain, biological, and transportation networks, multiple types of proximities exist. Multiplex networks, where nodes can have interactions in multiple views, have been proposed for accurately modeling such applications~\cite{main-ML}. 

While CS in multiplex networks recently has attracted much attention and several community models have been proposed, these models suffer from three main limitations. \textbf{(1)} Structure inflexibility refers to the problem that they are based on pre-defined patterns or rules, e.g., ML-LCD~\cite{ML-LCD}, ML \textbf{k}-core~\cite{ml-core-journal}, and FirmTruss~\cite{FirmTruss}. However, the structure of a community is flexible in nature, and it is nearly impossible to produce a high-quality community directly using the pre-defined rules. \textbf{(2)} These methods treat each relation type (aka view) equally for identifying the community, while many real-world multiplex networks may contain noisy/insignificant views~\cite{FirmCore}. Moreover, all nodes in the network might not have full information in all views, and so these noisy/insignificant views may be different for each vertex~\cite{deep_partial_ml, FirmCore}. \textbf{(3)} These methods only consider the structural properties of the network, while in real-world applications, networks often come naturally endowed with attributes associated with their vertices.

To mitigate the above limitations, we propose \model\!,  a query-driven graph convolutional network in multiplex networks that encodes the information and knowledge from both the query and graph. In each relation type, \model uses two encoders, \qencoder and \vencoder\!, to embed the local query-dependent structure and graph query-independent structure. To take advantage of complementary information of different views, \model uses an attention mechanism to incorporate information on different relation types. Finally, we use a feedforward neural network model to classify vertices. Given a query vertex set, this model classify vertices of the network as ``\textit{belonging to the community}''~or~not. 

\begin{figure*}
\centering
    \subfloat[][\centering The proposed framework]{{\includegraphics[width=0.57\textwidth]{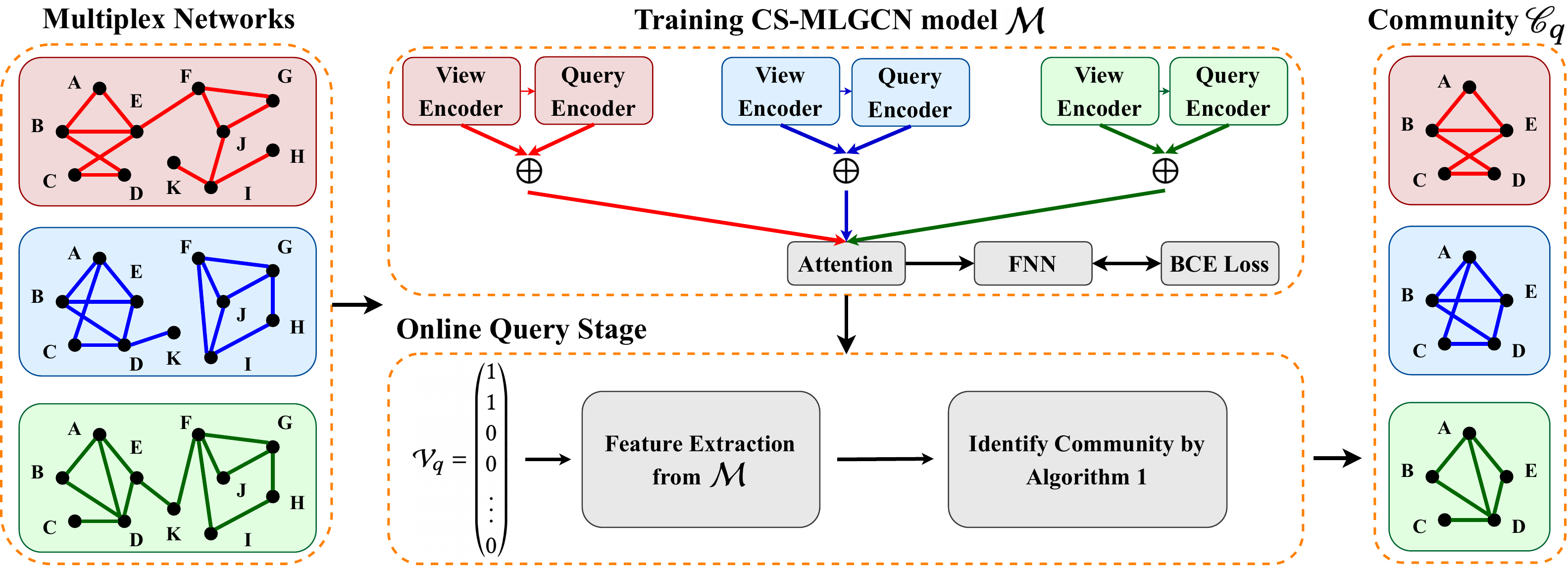} }}
    \hspace{2ex}
    \subfloat[][\centering Architecture of the CS-MLGCN model]{{\includegraphics[width=0.40\textwidth]{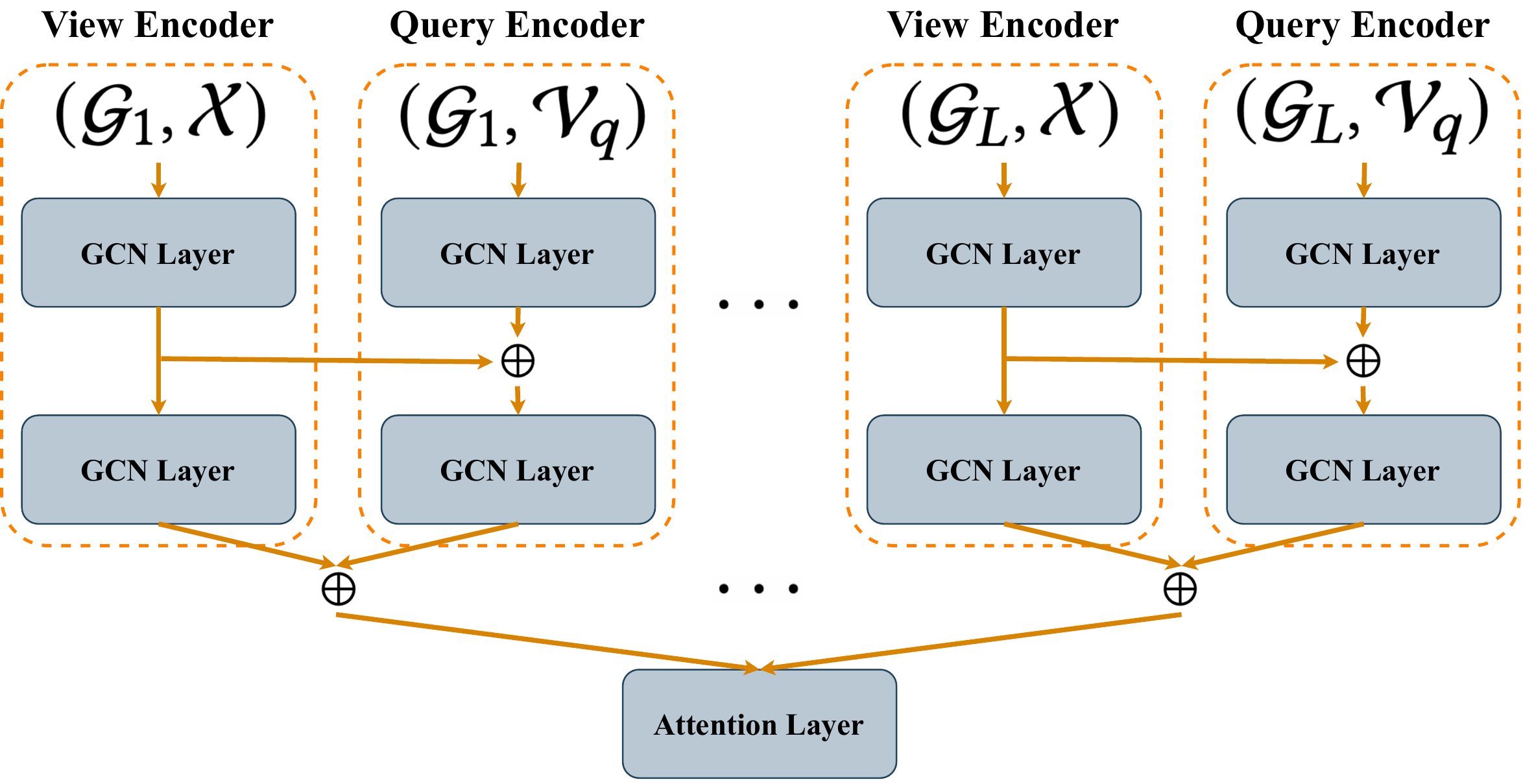} }}
    \vspace{-2ex}
    \caption{Framework and design of our proposed model.}
    \vspace{-2ex}
    \label{fig:model_framework}
\end{figure*}

\head{Motivations}
In multiplex networks, different views are complementary to each other. Accordingly, they provide more complex and richer information than simple networks and so can benefit typical applications of CS in simple networks~\cite{community_search_survey} (e.g., event organization, friend recommendation, etc.), delivering better solutions. However, next, we illustrate an exclusive application for community search in multiplex networks. 

Detecting functional systems in the brain is a fundamental task in neuroscience~\cite{functional_system_brain, functional_system_brain2}. A brain network is a graph in which vertices represent regions of the brain, and edges represent co-activation between regions. While a brain network generated from an individual subject might be noisy/insignificant, using brain networks from many subjects lets us identify important structures more accurately~\cite{FirmTruss, Brain_network_fmri}. A multiplex brain network is a multiplex graph in which each view represents the brain network of an individual. A CS method in multiplex networks (e.g., our method) can be used to identify the functional system of each brain region.

\head{Related Work}
Community search, which aims to find query-dependent communities in a graph, was introduced by Sozio and Gionis \cite{k-core-community}. Since then, various community models have been presented based on different pre-defined cohesive graph patterns~\cite{closest, TrussEquivalence, community2, clique-community, densest_community, attribute-community}. More recently, CS has also been investigated for more complex graphs, such as directed~\cite{community-directed, D-core_community_search}, geo-social~\cite{geo-social-community, road_social}, temporal~\cite{temporal-community}, multi-valued~\cite{multi-valued}, weighted~\cite{weighted-truss}, and labeled~\cite{butterfly_core} graphs. To address the structural inflexibility of these models, recently, graph neural network-based approaches have been introduced~\cite{GNN_CS, GNN_CS1}. However, all these approaches focus on networks with a single view while we study networks with multiple views.

In multiplex networks, Interdonato et al.~\cite{ML-LCD} design a greedy strategy to maximize the ratio of similarity between nodes inside and outside of the local community over all views. Galimberti et al. \cite{ml-core-journal} propose a community model based on the ML $\mathbf{k}$-core~\cite{azimi-etal}. Luo et al. \cite{ML-random-walk} design a random walk strategy to find local communities in multi-domain networks. Finally, recently, Behrouz et al.~\cite{FirmTruss} propose a community model based on the FirmTruss structure. However, all these models are based on pre-defined patterns or rules, which limits their applications.

\vspace{-2ex}
\section{Preliminaries}
In this paper, we follow the common definition of multiplex networks in the literature~\cite{multiplex_network_embedding, scalable_multiplex_embedding, FirmTruss}. We let $\GG = \{ \GG_r \}_{r = 1}^{L} = (\V, \E, \X)$ denote a multiplex graph, where $\GG_r = (\V, \E_r, \X)$ is a graph of the relation type $r$ (aka view), $\V$ is the set of nodes, $\E = \bigcup_{r = 1}^{L} \E_r$ is the set of edges, and $\X \in \mathbb{R}^{|\V| \times f}$ is a matrix that encodes node attributes information for nodes in $\V$. Given attribute matrix $\X$, $\X_v$ represents the attribute set of vertex $v \in \V$. Finally, the set of neighbors of node $v \in \V$ in type $r$ is denoted $N_r(v)$.

\vspace{-1ex}
\begin{problem}[Community Search in Multiplex Networks]\label{problem:community_search}
Given a multiplex network $\GG = (\V, \E, \X)$, and a vertex query set $\V_q \subseteq \V$, the problem of Community Search in Multiplex Networks (CSML) is to find the query-dependent community $\CC_q \subseteq \V$ that is connected and has a cohesive structure. 
\end{problem}
\vspace{-1ex}

In this paper, we formulate the above problem as a binary classification task. Given a set of query vertices $\V_q \subseteq \V$, we classify nodes in $\V$ into two classes: \textbf{1}: belonging to community $\CC_q$, \textbf{0}: otherwise. To this end, we use the one-hot vector $\cvec^{\text{out}}_q \in \{0, 1\}^{|\V|}$ to represent the output community $\CC_q$ by a model $\M$. That is, if ${\cvec^{\text{out}}_{q_v}} = 1$, vertex $v$ belongs to the predicted community $\CC_q$ by model~$\M$.

\section{Framework}
In this section, we introduce our proposed framework for CS in networks with multiple types of interactions. The overview of the framework is shown in Figure~\ref{fig:model_framework}(a). Our framework consists of two main stages. The first stage is to offline train a model $\M$ to learn to predict the membership of each vertex to the corresponding community of query vertices. The second stage is the online query stage. Once $\M$ is trained, given a query vertex set $\V_q$, we utilize the output of $\M$ for $\V_q$ and find a connected community that corresponds to $\V_q$. 

\vspace{-1ex}
\subsection{\model Model}
 The \model is based on the idea of graph convolutional networks and iteratively combines the embeddings of the local neighbourhood of each vertex. In multiplex networks, the neighbourhood of each node in each view is different. Moreover, the Vanilla GCN~\cite{vanilla_GCN} layer does not consider the type of connections in the training phase. One idea to overcome this challenge is to treat each view equally and incorporate the information of all views. However, in many real-world multiplex graphs, the data in some types of relations are noisy/insignificant~\cite{FirmCore, deep_partial_ml}, indicating lesser~importance.

 In our approach, we first use GCN layers to encode the information of a node in each view, called \vencoder\!. Second, to better capture the local query structure information, we use GCN layers and input query vector $\cvec_q$ as the features of nodes. We refer to this part as \qencoder\!. \qencoder enables query-centered structural propagation. That is, propagating from the query vertices to its neighborhood. Next, we take advantage of the attention mechanism and automatically learn the importance weights of each interaction type for each node. Based on these learned weights, we incorporate the type-specific embeddings of each node. Finally, we use a feedforward neural network (FNN) for classification. The overall architectures of \vencoder and \qencoder are shown in Figure~\ref{fig:model_framework}(b). Next, we explain each part in detail.

\head{View Encoder}
Given a relation type $r$, \vencoder encodes the global structure and vertex attributes of each view $\GG_r$, both of which are independent of query vertices. To this end, it employs the layer-wise forward propagation of GCN with a self feature modeling as:
\vspace{-2ex}
\begin{equation}\label{eq:propagatiion}
    {h_u^r}^{(\ell + 1)} \hspace{-0.5ex} = \Dr\left\{\sigma \left({h_u^r}^{(\ell)} {\W_{\text{s}}^r}^{(\ell + 1)}\hspace{-0.5ex} + \hspace{-2ex} \sum_{v \in N_r(u)} \hspace{-1ex} [\frac{{h_v^r}^{(\ell)}}{\sqrt{p^r_vp^r_u}} {\W^r}^{(\ell + 1)} \hspace{-0.5ex} + {\bb^r}^{(\ell + 1)}] \right)\right\},
\end{equation}
where in each view $\GG_r$, ${h^r_u}^{(\ell + 1)} \in \mathbb{R}^{d^{(\ell + 1)}_r}$ is the learned new features of node $u$ in the $(\ell + 1)$-th layer, ${h^r_u}^{(\ell)} \in \mathbb{R}^{d^{(\ell)}_r}$ is the hidden feature of $u$ in $\ell$-th layer, and ${\W_{\text{s}}^r}^{(\ell + 1)}, {\W^r}^{(\ell + 1)} \in \mathbb{R}^{d^{(\ell)}_r \times d^{(\ell + 1)}_r}$, and ${\bb^r}^{(\ell + 1)}~\in~\mathbb{R}^{d^{(\ell + 1)}_r}$ are trainable weights. $\sigma(.)$ is a nonlinearity, e.g., ReLU, and $\Dr(.)$ is the dropout method~\cite{dropout} to avoid overfitting. Given a relation type $r$ and a node $u \in \V$, $p^r_u$ denotes the degree of node $u$ in view $\GG_r$ plus one, i.e., $p^r_u = \mathrm{deg}_r(u) + 1$. The input feature of node $u$ in the first layer, ${h^r_u}^{(0)} \in \mathbb{R}^d$, is the normalized feature vector $\X_u$. This part lets us encode attribute and structural information and retrieve the query-independent~knowledge.

\head{Query Encoder}
Given a query vertex set $\V_q$, we first encode it to a one-hot vector $\cvec_q$, such that ${\cvec_q}_u = 1$ if $u \in \V_q$, and ${\cvec_q}_u = 0$ otherwise. Now, for each relation type like $r$, the type-specific propagation function for vertex $u\in\V$ is identical to Eq.~\ref{eq:propagatiion}. We denote query-dependent hidden
feature of $u \in \V$ in $\ell$-th layer by ${{h^r_Q}_u}^{(\ell)}$, and let ${\W_{Q}^r}^{(\ell + 1)}_{\text{s}}, {\W_Q^r}^{(\ell + 1)}$, and ${\bb_Q^r}^{(\ell + 1)}$ be trainable query-dependent weights. Contrary to \vencoder, here, the input feature of vertex $u$ in the first layer ${{h^r_Q}_u}^{(0)}$ is the one-hot query vector ${\cvec_q}_u$. As shown in Figure~\ref{fig:model_framework}(b), we aggregate the output of each layer in \vencoder and \qencoder and consider it as the input of the next layer of \qencoder\!. This approach lets us provide stable and robust knowledge about the query-independent structure and nodes' features of the graph for the \textit{Query Encoder}. Overall, \qencoder lets us provide an interface for query vertices and retrieve the local structure knowledge.

\head{Attention}
The main role of this attention mechanism is to incorporate the information of different views in a weighted manner. As we discussed in Sec.~\ref{sec:introduction}, the importance of each view for each node might be different, and so we cannot consider a single weight for each view. To this end, we suggest an attention mechanism that learns the importance of view $r$ for an arbitrary node $u\in \V$. Let $\zeta_u$ be the aggregated hidden feature of node $u\in \V$, and $\alpha^r_u$ indicate the importance of view $r$ for $u$, then:  
\vspace{-1ex}
\begin{equation}
    \zeta_u = \sum_{r = 1}^{L} \alpha^r_u \tilde{h}^r_u, 
    \vspace{-1ex}
\end{equation}
\vspace*{0ex}
where $\tilde{h}^r_u$ is the aggregation of the \textit{View Encoder}'s and \textit{Query Encoder}'s output for node $u$. Following the recent attention based models~\cite{multiplex_network_embedding, attention_main}, we use the softmax function to define the importance weight of view $r$ for node $u$:
\begin{equation}
    \alpha^r_u = \frac{\exp\left( \sigma\left( {\textbf{s}_r}^T \W^{r}_{\text{att}} \tilde{h}^r_u \right) \right)}{\sum_{k = 1}^{L} \exp\left(  \sigma\left( {\textbf{s}_k}^T \W^{k}_{\text{att}} \tilde{h}^k_u \right)  \right)},
\end{equation}
where $\mathbf{s}_r$ is the summary of view $r$, i.e. $\mathbf{s}_r = \sum_{u \in \V} \tilde{h}^r_u$, and $\W_{\text{att}}^r$ is a trainable weight matrix. In our experiments, we use $\tanh(.)$ as the activation function $\sigma(.)$.

\head{FNN Layer}
Finally, a feedforward neural network is used to classify nodes based on the~obtained~embedding~from~the~previous~part.

\head{Loss Function}
As we discussed, we formulate the CSML problem as a binary classification task. Given a query vertex set $\V_q$, let $\psi_q \in \mathbb{R}^{|\V|}$ be the output of model $\M$. Accordingly, ${\psi_q}_u \in [0, 1]$, and it represents the probability of $u$ being a member of $\CC_q$. Let ${\mathbf{y}_q}_u$ be the ground-truth label of $u$, based on the ground-truth community for $\V_q$, we utilize Binary Cross Entropy (BCE) as the loss function. Formally, given a query vertex set $\V_q \subseteq \V$, the query-dependent loss function $\mathcal{L}_q$ is defined as:
\begin{equation}
    \mathcal{L}_q = \frac{1}{|\V|} \sum_{u \in \V} - \left( {\mathbf{y}_{q}}_u \log({\psi_q}_u) + (1 - {\mathbf{y}_{q}}_u) \log(1 - {\psi_q}_u) \right).
\end{equation}

\vspace{-1ex}
\head{Training}
Given a set of query vertex sets $\mathcal{Q}_{\text{train}} = \{ q_1, q_2, \dots, q_n \}$, and their ground-truth communities $\CC_{\text{train}} = \{ \mathcal{C}_1, \dots, \mathcal{C}_n \}$, we first construct all query inputs as one-hot vectors. Then we repeatedly feed a query $q \in \mathcal{Q}_{\text{train}}$ into $\M$ and use $\M$'s output, $\psi_q$, to compute BCE loss and gradients of the model parameters.
With the updated parameters, iteratively, $\M$ goes to the next iteration and gets another query $q' \in \mathcal{Q}_{\text{train}}$ as input, until it passes all queries in the $\mathcal{Q}_{\text{train}}$. The overall loss function is the sum of all query-dependent loss functions, i.e.: 
\begin{equation}
    \mathcal{L} = \sum_{q \in \mathcal{Q}_{\text{train}}} \mathcal{L}_q
\end{equation}

\begin{algorithm}[t]
    \small
    \caption{Multiplex Community Identification}
    \label{alg:MCI}
    \begin{algorithmic}[1]
        \Require{A multiplex network $\GG = (\V, \E, \X)$, a query vertex set $\V_q$, a trained model $\M$, and a threshold $\eta$}
        \Ensure{A multiplex community $\CC_q$}
        \State $Q \leftarrow \V_q$; $\CC_q \leftarrow \V_q$;
        \State $\psi_q \leftarrow$ feed $\V_q$ into $\M$;
        \While{$Q$ is not empty}
            \State pick and remove a vertex $u$ from $Q$;
            \For{any relation type $r$}
                \For{$v \in N_r(u)$ and ${\mathbf{\psi}_q}_v \geq \eta$}
                    \State $Q \leftarrow Q \cup \{v\}$; $\CC_q \leftarrow \CC_q \cup \{v\}$;
                \EndFor
            \EndFor
        \EndWhile
        \Return $\CC_q$
    \end{algorithmic}
\end{algorithm}

\vspace{-1ex}
\subsection{Online Query for CSML}
In this section, we discuss how we can use the trained model $\M$, to process the online query $\V_q$ and identify the community $\CC_q$ without re-training $\M$.

\head{Community Identification Algorithm}
Given a query vertex set $\V_q$, we first construct its representative one-hot vector, and input it to model $\M$. Let $\psi_q \in \mathbb{R}^{|\V|}$ be the output of model $\M$, ${\psi_q}_u \in [0, 1]$ and represents the probability of $u$ being a member of $\CC_q$. We present the details of online query stage in Algorithm~\ref{alg:MCI}. Given a threshold $\eta \in [0, 1]$, we identify $\CC_q$ as the set of vertices like $u$ such that ${\psi_q}_u \geq \eta$. To ensure connectivity, we use Breadth-First Search (BFS) traversal starting from query vertices. 

\head{Time Complexity}
Algorithm~\ref{alg:MCI}'s time is dominated by GCN inferring, as the BFS traversal in this algorithm takes $\mathcal{O}(|\E|)$ time.

\section{Experiments}

\head{Setup}
We implement our methods using Python 3.7 and Pytorch. In each encoders, we build two layers with 128 neurons in the hidden layer, and train the model 200 iterations with a learning rate of 0.001. Unless stated otherwise, in each layer we employ ReLU as activation function, and dropout rate of 0.5. For large datasets, we adopt the method used in~\cite{GNN_CS, GNN_CS1} and select nodes in 2-hop neighbors of query nodes as the candidate subgraph for each query. We then train the model on these~small~subgraphs~and~predict~communities.

\head{Datasets}
We perform experiments on seven real networks ~\cite{FirmTruss, amazon_datset, Noordin-dataset, RM, brain_dataset, DKPol} containing ground-truth communities, whose domains cover social, co-authorship, brain, and co-purchasing networks. Their main characteristics are summarized in Table~\ref{tab:datastat}. It is noteworthy that ground-truth communities in the Brain dataset are functional systems of the human brain. Accordingly, its results can demonstrate the effectiveness of CS methods in detecting functional systems of the brain. For each dataset, we consider $\min\{|\V|, 350\}$ query-community pairs and split them into training, validation, and test data with the ratio of 150:100:100. Training data is used to train the model, validation data is used to select the best hyperparameters, and test data is used to measure the performance~of~all~methods.

\begin{center}
\begin{table} [tpb!]
 \caption{Network Statistics}
 \vspace{-2ex}
    \resizebox{0.40\textwidth}{!} {
\begin{tabular}{l | c c c c c c c}
 \toprule
  {Dataset} & {AUCS} &  {Terrorist} & {RM} &  {DKPol} & {Brain} & {DBLP} & {Amazon}\\
 \midrule 
 \midrule
    $|\V|$  &   61     & 79      & 91     & 490     &    190& 513K & 410K  \\
    $|\E|$     &    620   &  2.2K    &  14K   & 20K   & 570K & 1M & 8.1M  \\
    \#Views     &    5   &   14    &  10     &  3   & 20    & 10 & 4 \\
 \bottomrule
\end{tabular}
}
 \label{tab:datastat}
 \vspace{-3ex}
\end{table}
\end{center}

\vspace{-5ex}
\head{Baselines}
We compare our approach with the state-of-the-art CS methods in multiplex networks. FTCS~\cite{FirmTruss} finds the connected FirmTruss containing query nodes with the minimum diameter.  ML $\mathbf{k}$-core~\cite{ml-core-journal} uses an objective function to automatically choose a subset of layers and then finds a subgraph such that the minimum of per-layer minimum degrees, across selected layers, is maximized. ML-LCD~\cite{ML-LCD} uses a greedy strategy to maximize the ratio of similarity between nodes inside and outside of the local community. RWM~\cite{ML-random-walk} sends random walkers in each view to obtain the local proximity w.r.t. the query nodes and returns the set of nodes with the smallest conductance. For baselines, we tune the parameters according to their original papers and report the best results.

\begin{figure}
    \centering
    \hspace{-3ex}
    \includegraphics[width=0.48\textwidth]{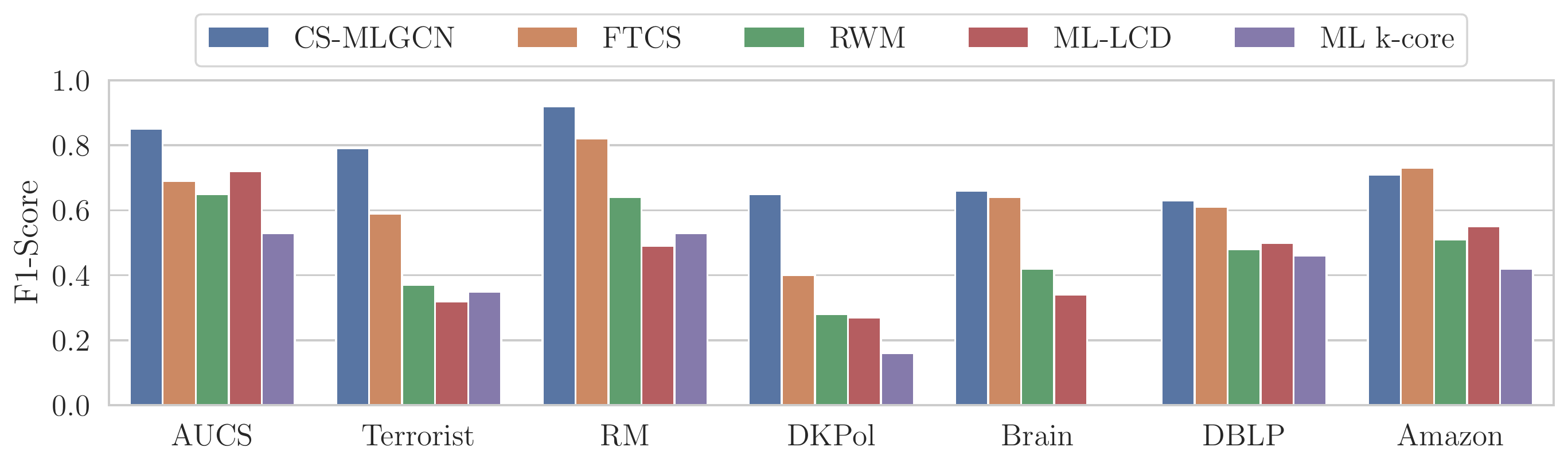}
    \vspace{-2.5ex}
    \caption{Quality evaluation.}
    % \vspace{-1ex}
    \label{fig:F1-score}
\end{figure}

% \vspace{-1ex}
\head{Quality Evaluation}
Figure~\ref{fig:F1-score} reports the average F1-scores of our method and baselines on datasets with the ground-truth community. ML \textbf{k}-core result on the Brain dataset is excluded as it did not terminate before a day. \model achieves the best or competitive results on all networks against baselines. The reason is two-fold. First, while the structure of a community is flexible in nature, these methods consider a pre-defined subgraph pattern, which causes inaccuracy in communities without such a pre-defined pattern. Second, in some datasets like RM and Terrorist, some views are noisy/insignificant for each vertex. Our approach by employing attention mechanism considers different importance weights for each relation type, and can detect important views for a vertex.

\begin{figure}
    \centering
    \hspace*{-1ex}
    \includegraphics[width=0.49\textwidth]{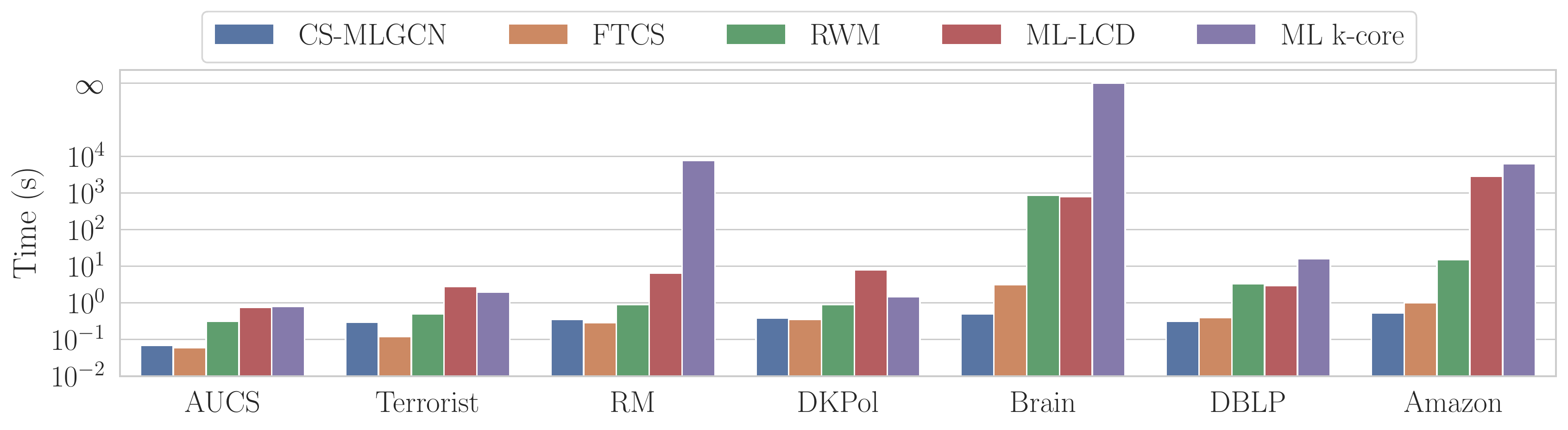}
    \vspace{-5ex}
    \caption{Efficiency Evaluation.}
    \vspace{-5ex}
    \label{fig:Runing_time}
\end{figure}

\head{Query Efficiency}
We evaluate the query efficiency of \model in the test set. 
Figure~\ref{fig:Runing_time} reports the average query processing time of all methods for 100 test queries. \model achieves a better or comparable query time compared to baselines. It is noteworthy that this result is achieved by our method, while it outperforms baselines in quality evaluation.

\begin{figure}
    \centering
    \hspace*{-4ex}
    \subfloat[][\centering Effect of $\eta$]{{\includegraphics[width=0.16\textwidth]{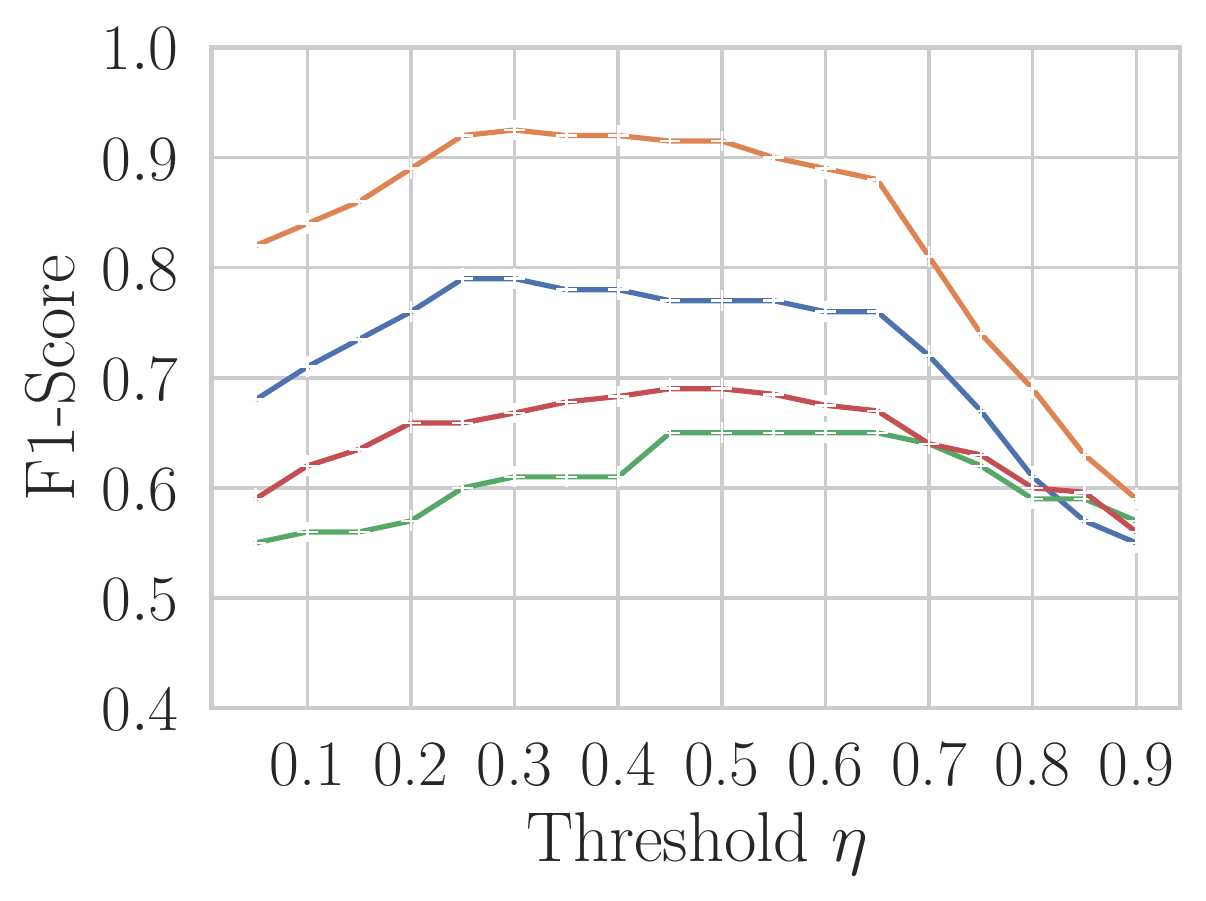} }}
    \subfloat[][\centering Effect of \#Epochs]{{\includegraphics[width=0.20\textwidth]{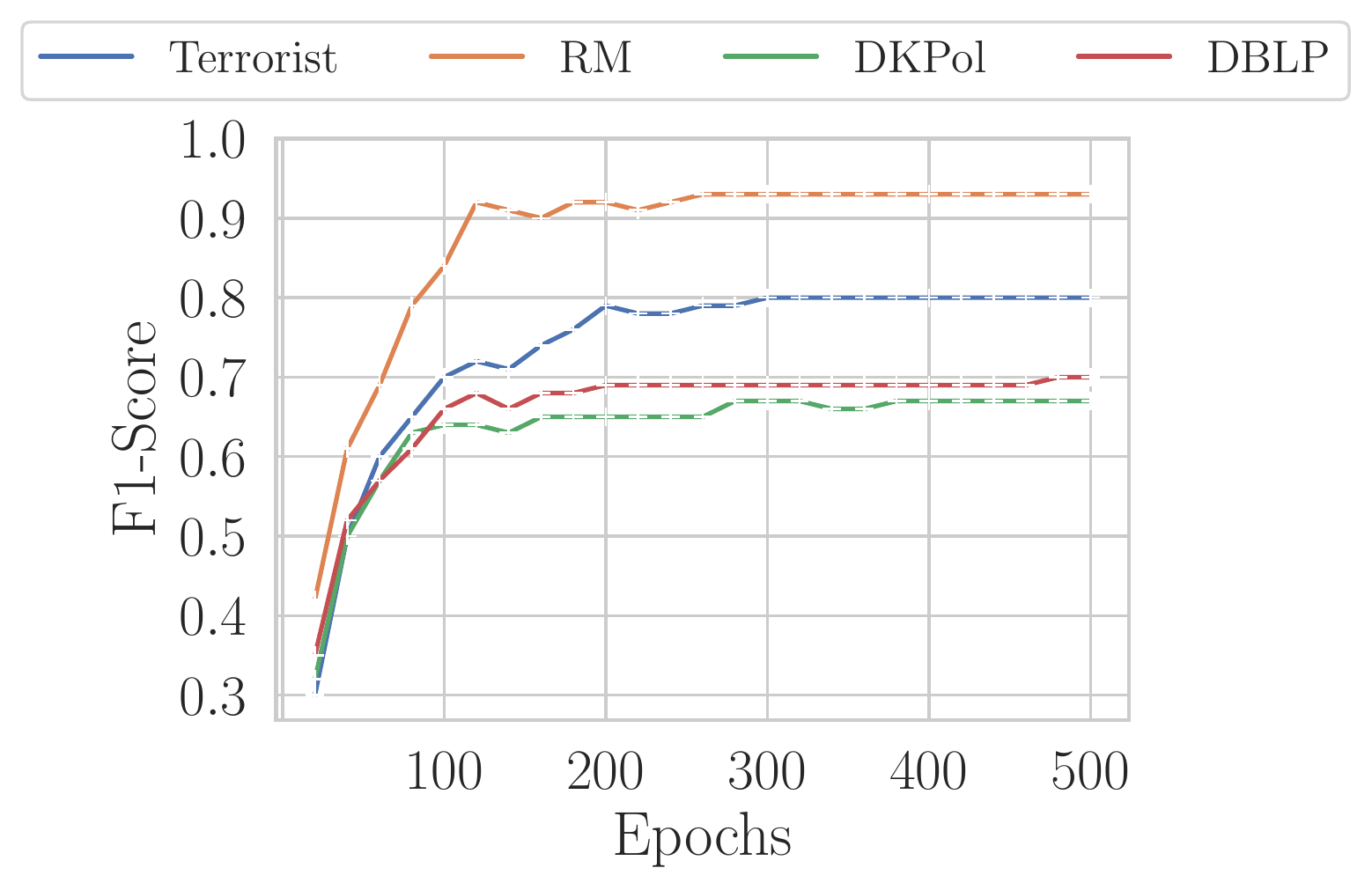} }}
    \hspace*{-2ex}
    \subfloat[][\centering Effect of Dropout Rate]{{\includegraphics[width=0.16\textwidth]{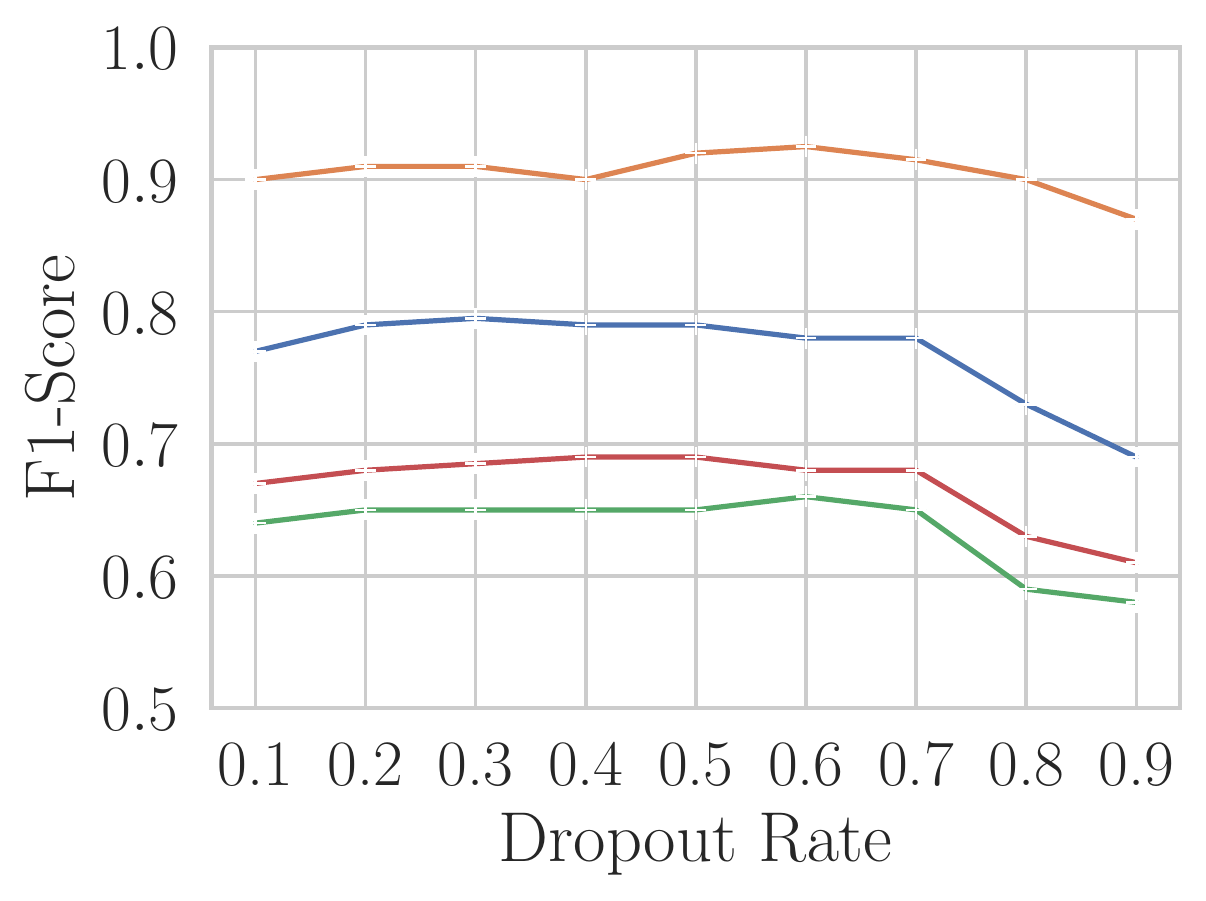} }}
    \vspace{-1.5ex}
    \caption{Parameter Sensitivity Evaluation.}
    \label{fig:ablation}
    % \vspace{-1ex}
\end{figure}

\head{Ablation Study}
In this section, we report the ablation studies of \model, i.e., the sensitivity evaluation of the parameter $\eta$, the epoch number, and dropout rate in the CSML problem. Parameter $\eta$ is used to translate the model output vector to the community vertex set. Figure~\ref{fig:ablation}(a) reports the F1-score for different values of $\eta$. While very large and very small value of $\eta$ affects the performance, for other values of $\eta$, the performance is almost stable. Figure~\ref{fig:ablation}(b) shows the effect of \#epochs on the performance. Here, we vary \#epochs from 0 to 500 and report the average F1-score of the test queries. \model achieves good results within 100 epochs. Finally, Figure~\ref{fig:ablation}(c) shows the effect of dropout rate. We observe that the performance is stable when the dropout rate is $\leq 0.7$.

\section{Conclusion}
In this paper, we propose \model\!, a query-driven GCN-based approach for CS in multiplex networks. For each view of the network, \model employs two encoders to encode the local query-dependent structure and global query-independent graph structure. To let different views collaborate with each other and incorporate their information, we proposed an attention mechanism that automatically learns the importance weights of views. Experiments on real-world networks demonstrate that the \model model outperforms previous approaches.

\everypar{\looseness=-1}

\bibliographystyle{ACM-Reference-Format}
\balance
\bibliography{main}

% \newpage

% $$\mathcal{M}$$

% $$\CC_q$$

% $$\mathcal{V}_q = \begin{pmatrix} 1\\ 1 \\ 0 \\ 0  \\ \vdots \\ 0\end{pmatrix}$$

% $$(\GG_1, \X)$$
% $$(\GG_1, \mathcal{V}_q)$$

% $$(\GG_L, \X)$$
% $$(\GG_L, \mathcal{V}_q)$$

% $$\oplus$$

% $$\dots$$

\end{document}